\documentclass[conference]{IEEEtran}
\IEEEoverridecommandlockouts
\usepackage{cite}
\usepackage{amsmath,amssymb,amsfonts}
\usepackage{algorithmic}
\usepackage{graphicx}
\usepackage{textcomp}
\usepackage{xcolor}

\usepackage[font=small,belowskip=-4pt,aboveskip=4pt]{caption}

\usepackage{subfiles}
\usepackage{mathtools}
\usepackage{comment}
\usepackage{subfig}
\usepackage{booktabs}
\usepackage{wrapfig}
\usepackage{multirow}
\usepackage{dashrule}
\usepackage{hyperref}

\newcommand{\Sys}{SmartON\xspace}

\usepackage[linesnumbered,ruled]{algorithm2e}
\SetKw{KwAnd}{and}
\SetKw{KwAlgo}{Algorithm}
\def\BibTeX{{\rm B\kern-.05em{\sc i\kern-.025em b}\kern-.08em
    T\kern-.1667em\lower.7ex\hbox{E}\kern-.125emX}}
    
\begin{document}

\title{\Sys: Just-in-Time Active Event Detection \\on Energy Harvesting Systems}

\author{\IEEEauthorblockN{Yubo Luo}
\IEEEauthorblockA{Department of Computer Science \\
UNC at Chapel Hill\\
yubo@cs.unc.edu}
\and
\IEEEauthorblockN{Shahriar Nirjon}
\IEEEauthorblockA{Department of Computer Science \\
UNC at Chapel Hill\\
nirjon@cs.unc.edu}

}

\maketitle

\begin{abstract}
  We propose \Sys, a batteryless system that learns to wake up proactively at the right moment in order to detect events of interest. It does so by adapting the duty cycle to match the distribution of event arrival times under the constraints of harvested energy. While existing energy harvesting systems either wake up periodically at a fixed rate to sense and process the data, or wake up only in accordance with the availability of the energy source, \Sys\ employs a three-phase learning framework to learn the energy harvesting pattern as well as the pattern of events at run-time, and uses that knowledge to wake itself up when events are most likely to occur. The three-phase learning framework enables rapid adaptation to environmental changes in both short and long terms. 
  Being able to remain asleep more often than a CTID (charging-then-immediate-discharging) wake-up system and adapt to the event pattern, \Sys\ is able to reduce energy waste, increase energy efficiency, and capture more events. To realize \Sys\ we have developed a dedicated hardware platform whose power management module activates capacitors on-the-fly to dynamically increase its storage capacitance.
  We conduct both simulation-driven and real-system experiments to demonstrate that \Sys\ captures 1X--7X more events and is 8X--17X more energy-efficient than a CTID system. All source code and hardware design files are open-sourced\footnote{https://github.com/YuboLuo/SmartON2021}.
\end{abstract}

\section{Introduction}

Current IoT world is dominated by battery-powered edge devices which are bulky and unsustainable during their lifetime.
Energy harvesting is one of the ways that can lead us to sustainable IoT where edge devices are powered solely by harvesting energy from the environment.

Event detection is a common task in IoT, including those that employ energy harvesting on edge devices. However, even with the advances in intermittent computing systems ~\cite{dewdrop,tragedy,capybara,lee2019intermittent} which improve energy efficiency either by dynamically changing turn-on voltage threshold or by automatically reconfiguring the storage capacitor, current energy harvesting systems are not optimal at event detection tasks when the energy harvesting and the arrival of events of interest are not correlated. For ease of discussion, we can broadly categorize event detection systems into two types --- \textit{passive} and \textit{active} event detection systems. Passive event detection works like an interrupt-enabled system, where an external trigger signal interrupts the system and wakes it up. Active event detection, on the other hand, works like a polling system, where the system actively wakes itself up according to a schedule. Since polling systems generally consume more energy than interrupt-based ones, enabling active event detection on energy harvesting systems is a challenging feat due to the intermittent, uneven, and scarce energy harvesting sources.

Existing works~\cite{campbell2016,campbell2014energy,campbell2014water,monjolo2013,contexttrigger,trinity2013,trinity2018,doubledip,stepcounter} that deal with event detection on energy harvesting systems are generally passive event detectors.  
For example, ~\cite{campbell2014energy} proposes a batteryless system that detects door-opening events, where the action of opening the door itself vibrates a piezoelectric sensor to generate the trigger signal that wakes the microcontroller (MCU) up. 
In the real world, however, there are many applications, such as air quality monitoring~\cite{contexttrigger} and environmental noise detection where the target events can only be captured through active sensing. There exists a body of works that allow a system to adapt the duty cycle of energy harvesting sensors~\cite{ACES,ACES_workshop,rlman,roy2018rl} which can be classified as active event detectors. However, these systems only take into account the harvestable energy from the energy source to determine the duty cycle. 
Since in an active event detection system, the arrival of events is generally independent of the energy source, relying solely on the energy harvesting pattern while ignoring the probability of event arrivals is not an effective strategy.  

In this paper, we propose \Sys\ which is the first standalone batteryless active event detection system that considers both the event arrival pattern as well as the harvested energy to determine \emph{when the system should wake up} and \emph{what the duty cycle should be}. 
The technical contributions of this work are two-fold:            
 
$\bullet$ We design a dedicated power management circuit which activates and deactivates individual capacitors in a custom-built capacitor array on-demand and on-the-fly, in order to let the system borrow harvested energy from previous charging cycles that is worth saving for later use instead of immediate discharging. This feature is particularly important in active event detection systems where events and harvestable energy are not aligned in time.     

$\bullet$ We propose a three-phase learning framework where Phase-1 learns the right moment to wake the system up, Phase-2 employs reinforcement learning to determine a suitable frequency for duty-cycling when the system is awake, and Phase-3 controls the system's ability to adapt at run-time. To expedite Phase-2, which is the most critical and time-consuming step, we partition the reinforcement learning process in such a manner that the learning process can be partially converged and become exploitable immediately without waiting for the whole learning process is finished.

We implement and evaluate a real system that executes audio and visual event detection tasks and runs on solar and RF energy harvesters separately. 
Our experimental results show that \Sys\ captures 1X--7X more events and is 8X--17X more energy-efficient than a CTID system.

\section{Just-in-Time Event Detection}

The goal of just-in-time event detection is to make the best use of harvested energy by waking the system up only when events are most likely to occur. In this section, we discuss the motivation and challenges toward achieving this goal.

\subsection{Motivation}

Energy harvesting systems typically have predefined turn-on and turn-off voltage thresholds to activate and deactivate the system. So, the system periodically wakes up at each capacitor charging cycle. We refer to these systems as \emph{charging-then-immediate-discharging} systems, denoted as \textbf{CTID}, where once the capacitor is charged to the turn-on threshold, it starts to discharge immediately and continues to discharge until reaching the turn-off threshold. Although some systems dynamically change the thresholds according to the complexity of the next task~\cite{capybara, dewdrop}, they still belong to the category of periodic charging-then-immediate-discharging systems.

\begin{figure}[!tb]
    \setlength{\belowcaptionskip}{-18pt}
    \includegraphics[width=0.47\textwidth]{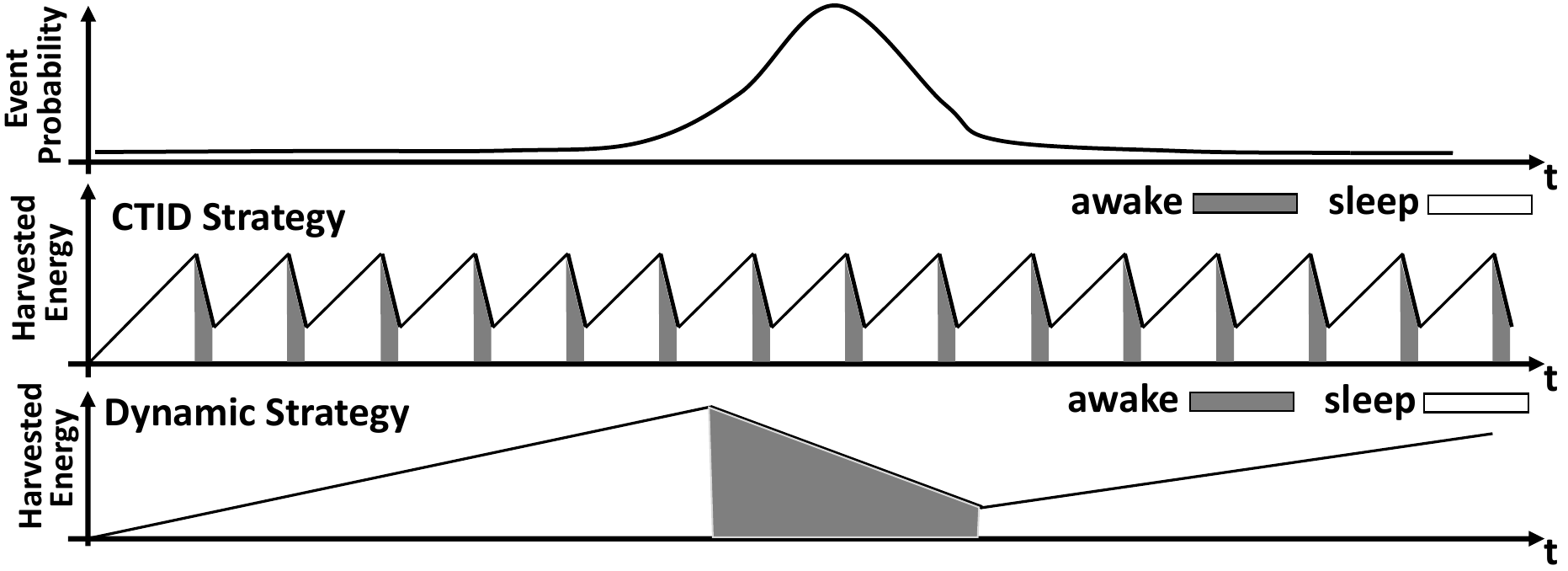}
    \vspace{-0.3\baselineskip}
    \caption{A comparison between CTID and dynamic strategy. A CTID strategy can only help detect a small fraction of total events, whereas a dynamic strategy helps detect more events by waking up when events are more likely to happen.}
    \label{fig:intro_threeStrategies} 
\end{figure}

The key to enabling just-in-time active event detection on energy harvesting systems is to ensure that the system knows when the events are more likely to occur, so that it could wake up more frequently during that interval, while remaining in a low-power mode at other times. 
Additionally, when events are unlikely to occur, the system should store the surplus energy (i.e., harvested energy not used in low-power mode) in a dedicated capacitor array to compensate for the excessive energy consumption due to frequent wake ups when events are happening. With this just-in-time discharge strategy, a system can borrow harvested energy from previous charging cycles.

In Fig.~\ref{fig:intro_threeStrategies}, we show a simple bell-shaped probability distribution for event arrivals (as an example) and two different wake-up strategies for an energy harvesting system. In the \emph{CTID} strategy, the system charges and then immediately discharges, whereas in the \emph{dynamic} strategy, the system charges until it expects to encounter the events. The active time (i.e., denoted as awake) of both systems is shaded in gray. We observe that although both systems remain active for the same amount of time over the timeline, unlike the CTID strategy, the dynamic strategy enables the system to spend most of its active time when events are happening.

\begin{figure}[!tb]
    \setlength{\belowcaptionskip}{-18pt}
    \includegraphics[width=0.48\textwidth]{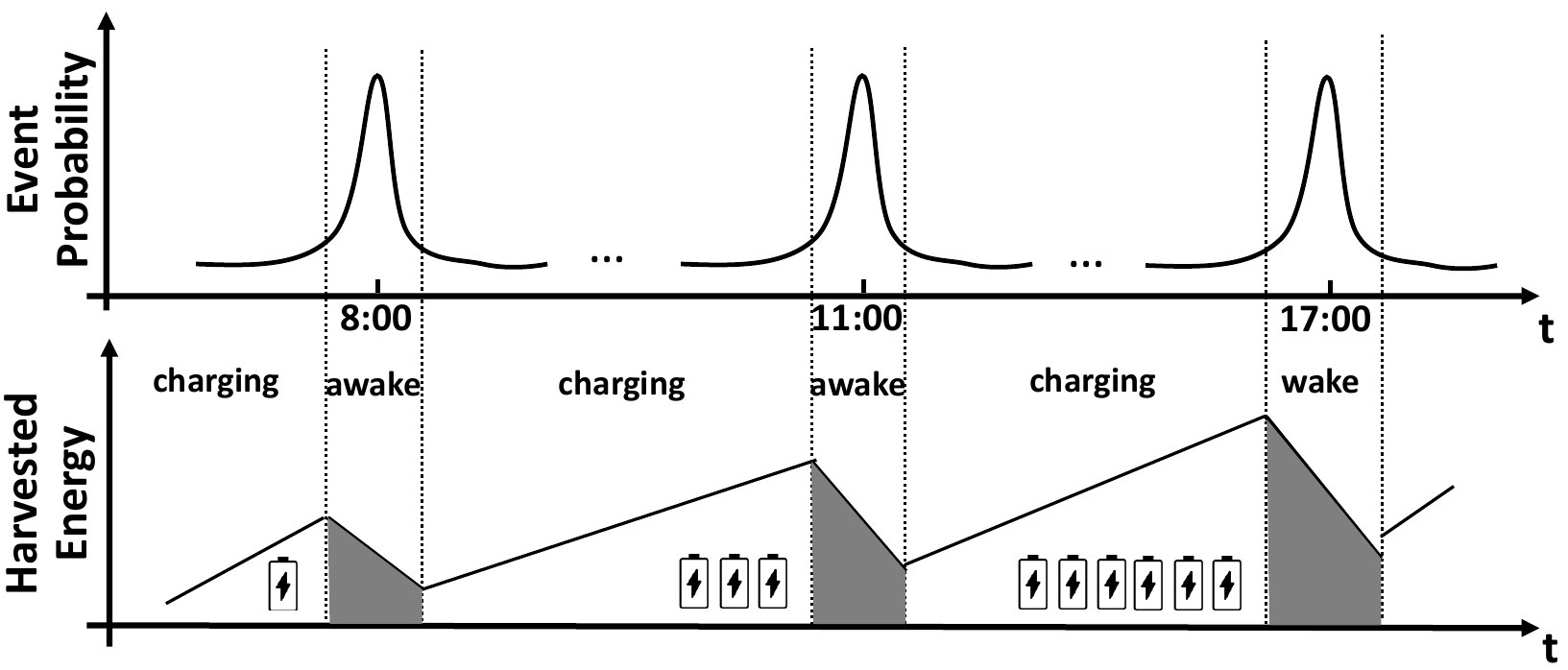}
    \caption{Capacitors are added to the system on-the-fly. }
    \label{fig:varying-charging-capacity}

\end{figure} 

\begin{figure*}[t] 
    \setlength{\belowcaptionskip}{-15pt}
    \centering
    \subfloat[System architecture.] 
    {
        \includegraphics[width = 0.25\textwidth]{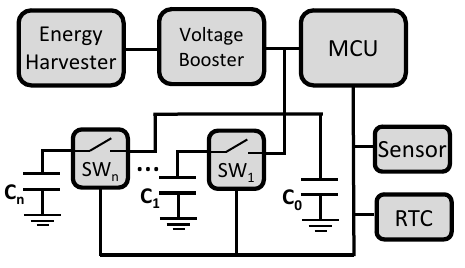}
    }
    \subfloat[Prototype board.]
    {
        \includegraphics[width = 0.20\textwidth]{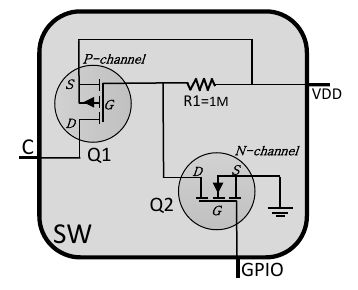}
    }
    \subfloat[Switch Unit.]
    {    
    \includegraphics[width = 0.22\textwidth]{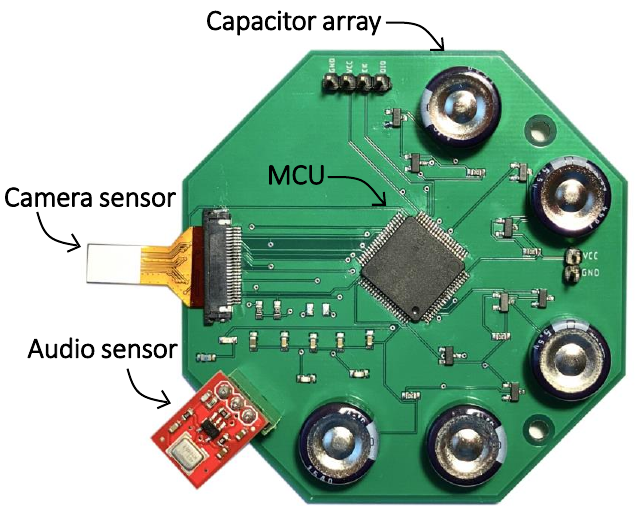}
    }
    \subfloat[Multiphase transfer diagram.]
    {    
    \includegraphics[width = 0.24\textwidth]{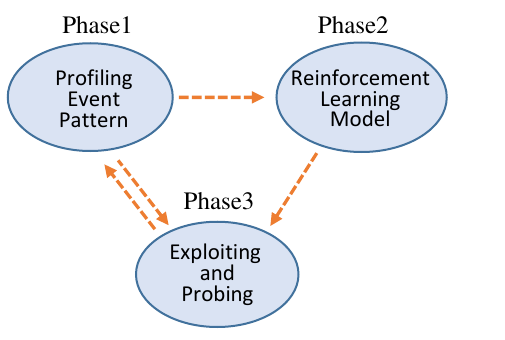}
    }

    \caption{\footnotesize{Overview of \Sys.}} 
    \label{fig:sys_desgin} 
\end{figure*}

\subsection{Challenge 1 -- Hardware Platform}

A major challenge to enable active event detection on an energy harvesting system is the saturation of the storage capacitor.

We envision a dedicated power management hardware for energy harvesting systems which will allow the system to incrementally activate capacitors from an array of capacitors when the currently activated capacitors collectively reach the charging saturation. Thus, the system will be able to store harvested energy without worrying for charging saturation, which will make it possible for the system to borrow unused energy from previous charging cycles.

Fig.~\ref{fig:varying-charging-capacity} illustrates a scenario where a system harvests solar energy and the probability distribution of event occurrence has three peaks. 
Assuming that the sun rises at 6 AM, the amount of harvestable energy before the first peak is low, as the time period is short and sunlight is weak during these hours. 
During noon and afternoon hours, more energy is harvested because of the longer charging period and stronger sunlight. The envisioned system activates more capacitors on-the-fly to avoid charging saturation and stores the harvested energy. 

Note that in this example, even if we use a single capacitor which is big enough to avoid charging saturation during the whole day, it can miss the morning events due to its longer time to charge to a voltage level which is high enough to drive the sensor~\cite{smith2013wirelessly}. On the other hand, if we use a small capacitor which can operate the sensor quickly, it may suffer from charging saturation in the afternoon. 
A dedicated hardware scheme solves this dilemma.

\subsection{Challenge 2 -- Learning Strategy}

The success of a just-in-time event detection system largely depends on its ability to predict the distribution of event arrival pattern accurately and being able to control its wake up frequency accordingly. 

For some applications, the predictability of events maybe high (e.g., traffic on a specific street on a specific time of the day), but generally, the parameters of the distribution in most real-world scenarios will vary and thus cannot be learned offline. Hence, a system must learn to estimate the event distribution -- e.g., the shape of the distribution and the positions of the onsets and peaks -- using online learning algorithms. The system must also determine the wake up frequency, given the probability of events and the amount of harvested energy at each time point.

Given the requirements as above, a reinforcement learning approach is a natural fit to the problem. A straightforward formulation of that would be to define a state space consisting of all possible combinations of event arrival probabilities and harvested energy, and to learn the corresponding wake up frequency as the actions. However, the number of (state, action) pairs in this straightforward formulation will be too large, and the learning process may take forever to converge.

In order to expedite the convergence, we must look for an alternative and efficient way to apply reinforcement learning. A key to reduce the problem is to break it down into multiple smaller learning problems that require learning only once.

\section{Overview of \Sys}

\subsection{\Sys\ Hardware} 
\label{hardware}

We design a custom hardware platform that primarily consists of: an energy harvester, a voltage booster, a capacitor array, a microcontroller, and sensors. Fig.~\ref{fig:sys_desgin}(a) shows a high-level schematic diagram of \Sys\ hardware architecture, our customized prototype board and the detailed design of switch unit.

$\bullet$ \emph{Capacitor Array}: \Sys\ implements an array of capacitors where each capacitor (except for one -- which is always active) is individually controlled by a custom-designed switch. These switches remain in their off position in the beginning. When the currently activated capacitors collectively reach their saturation, the MCU activates a new capacitor from the array using one of its General Purpose Input/Output (GPIO) pins.      

Due to the non-linearity of capacitor charging, the mechanism of activating capacitors on-the-fly in \Sys\ has higher charging efficiency than any hardware design that uses a single capacitor. Activating a new capacitor causes a voltage drop of the entire capacitor array. By setting a threshold over which a new capacitor will be activated, \Sys\ makes it possible for the capacitor array to be always charging at a low voltage level and thus it achieves a higher charging efficiency.

$\bullet$ \emph{Switch Unit}:
The detailed design of switch unit is shown in Fig. \ref{fig:sys_desgin}(b). Each switch contains a P-channel and an N-channel enhancement-type metal–oxide–semiconductor field-effect transistor (MOSFET). Both MOSFETs are normally off. 
The N-channel MOSFET requires a positive $V_{GS}$ to switch on, whereas the P-channel MOSFET requires a negative $V_{GS}$ to switch on. 
By default, the GPIO pin is set to LOW and the two MOSFETs are off, so that the capacitor is disconnected from the VDD. 
When the MCU wants to activate a capacitor, it changes the GPIO pin to HIGH -- which turns on the N-channel MOSFET and grounds its drain (D). 
On the other hand, the gate (G) of P-channel MOSFET will also be grounded and P-channel MOSFET now has a negative $V_{GS}$, and thus it is also turned on -- which connects the capacitor to the VDD.

This design is more energy efficient and simpler (i.e., having less number of electronic components) than the state-of-the-art dynamically expandable capacitor array~\cite{capybara}. ~\cite{capybara} is a charging-then-immediate-discharging system where the MCU remains completely powered off when the capacitor is charging. Therefore, their design needs extra circuits to maintain the state of the switch when the MCU is off. 
In contrast, the MCU in \Sys\ is in the low-power mode when it is charging thus the switch state is locked and we do not need extra circuits to maintain the switch state.

$\bullet$ \emph{Other Components}: We use a TI MSP430FR5994 microcontroller and an LTC3105 voltage booster. 
Capacitors are activated in the ascending order of their capacitance to make sure later added capacitors can bring down the voltage of the entire array to a lower level.
An internal timer is maintained to keep track of the time during this period. When the system is completely powered off, e.g., due to intermittent energy, an external real-time clock (RTC) module is used to keep track of the world time. Using the RTC simplifies the hardware design. However, we can replace this with a completely batteryless time keeper such as Botoks~\cite{botoks}. For sensors, we have tested the system with both camera and microphone sensors.

\subsection{\Sys\ Learning Framework}

The goal of \Sys\ is to remain in the low-power mode for most of the system lifetime when events are unlikely to happen, and wake up at the right time and operate at the right frequency as events happen. 
To enable this, we implement a three-phase learning framework that divides the learning problem into three phases where each phase focuses on a particular aspect of the learning. Fig.~\ref{fig:sys_desgin}(d) shows the three phases of learning framework:

$\bullet$ \emph{Phase-1}: The system uses a profiling algorithm to achieve two goals: (1) to learn the right moment for system to wake up, and (2) to learn the shape of the event distribution. Once Phase-1 finishes, it either moves to Phase-2 or directly moves to Phase-3. Being able to short-circuiting Phase-2, the system saves a significant amount of time and energy that an alternative monolithic learning framework would incur.    

$\bullet$ \emph{Phase-2}: The system employs reinforcement learning to determine the wake up frequency, given the amount of harvested energy of that moment and the shape of the event distribution that it is about to experience. By exploiting the repeated patterns in energy harvesting as well as event arrivals, this step reduces the state space of reinforcement learning and converges orders of magnitude faster than a straight-forward adoption of reinforcement learning that does not exploit the repeated subproblems. Once the reinforcement learning converges, it switches to Phase-3. 

$\bullet$ \emph{Phase-3}: The system \emph{exploits} (i.e., keeps using the learning model) the converged reinforcement learning model while periodically probing to detect if the event pattern has changed. If so, the system switches to Phase-1. Note that if Phase-1 detects that the shape of the distribution remains the same but only the distribution has drifted, the system can directly go back to Phase-3 as the reinforcement learning depends on the shape of the distribution, not its positional changes.

The application of reinforcement learning in \Sys\ is significantly different from previous works~\cite{ACES,ACES_workshop,rlman,roy2018rl} which also use reinforcement learning to adapt the duty cycle of the system but have a different goal -- which is to ensure the continuity of the sensing without caring for the energy waste when events are unlikely to occur. These works do not consider event distributions, and do not exploit the repeated subproblems in the learning process to expedite its convergence.

\section{Algorithmic Details}
\label{section:approach}

This section provides the algorithmic details of each of the three phases of the learning framework. 

\subsection{Phase-1 -- Profiling Event Pattern} 

The purpose of the profiling algorithm --- which is used in the Phase-1 of \Sys\ learning framework --- is to learn the event distribution over a time period. We discretize the period into fixed-sized time slots for the profiling purpose. For example, if the profiling period is a day, each slot can be an hour of the day.

\begin{figure}[!tb] 
    \setlength{\belowcaptionskip}{-15pt}
    \centering
    \includegraphics[width = 0.5\textwidth]{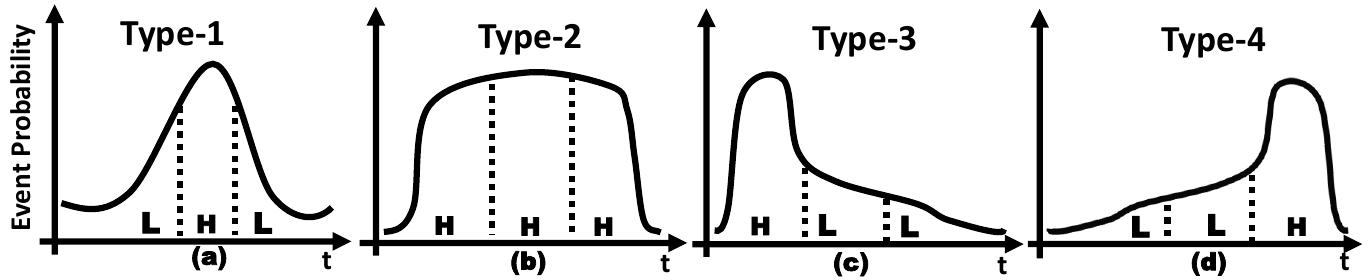} 
    \caption{Examples of different shaped peaks in the event distribution.}
    \label{fig:Algo_FourEvetTypes} 
\end{figure}

The profiling happens in real-time. During profiling, the system wakes up at each time slot at a high frequency and records the number of events it catches at each time slot.
In this way, the profiling phase reveals the time slots during when the system should be awake. Additionally, the number of caught events in each slot shows the likeliness of event happening in that slot. It reveals the shape of the distribution -- which we model as a superposition of a predefined set up \emph{shapes}. In Fig. \ref{fig:Algo_FourEvetTypes}, we show four example shapes. Each shape requires a different wake-up strategies to maximize the number of caught events. For example, for an event pattern that contains a shape like Fig.~\ref{fig:Algo_FourEvetTypes}(a), the system should allocate most of its harvested energy to the middle slot as events are more likely to occur in the middle. For shapes like Fig.~\ref{fig:Algo_FourEvetTypes}(b), energy should be evenly spent over the slots. For shapes like Fig.~\ref{fig:Algo_FourEvetTypes}(c) and Fig.~\ref{fig:Algo_FourEvetTypes}(d), most of the energy should be spent in the first or the last slot, respectively. 

\begin{figure}[!tb] 
    \setlength{\belowcaptionskip}{-18pt}
    \centering
    \includegraphics[width = 0.5\textwidth]{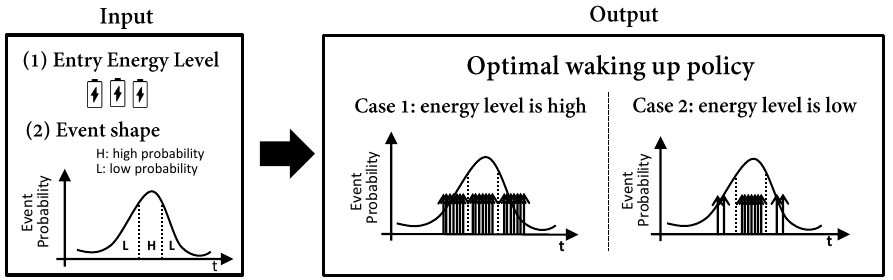} 
    \caption{The workflow of Phase-2. Given the inputs, including the storage energy level and the type of the event shape, Phase-2 should output an optimal wake-up strategy. For example, if the energy level is high, the system should wake up with high frequency at all times; if the energy level is low, the system should only wake up with high frequency in states where events are most likely to happen. }
    \label{fig:optimalwakinguppolicy} 
\end{figure}

Note that although we assume a predefined set of shapes that constitutes the event distribution, the exact value of the wake-up frequency is learned online in the Phase-2 of the proposed learning framework. This is because, the value of the harvested energy at the onset of each shape dictates the wake up strategy for the shape -- which is generally unknown. This is illustrated by Fig.~\ref{fig:optimalwakinguppolicy}.

\begin{algorithm}[!t]

\small
 visited[state.n] = \{0\}\;
 \While{not converged}{
     \For{$step\leftarrow 1$ \KwTo $state.n$}{
     
        check the current energy level of the storage capacitor array;
        \uIf{there is enough energy \KwAnd visited[step]==0}{
            profile this state with high wake-up frequency\;
            visited[step]=1\;
        }
        \uElse
        {remain asleep\;}
     }
 }
 \caption{Profiling Event Pattern}\label{algo:algorithm1}
\end{algorithm}

Algorithm~\ref{algo:algorithm1} shows the pseudocode of our profiling algorithm in Phase-1.  
The algorithm keeps track of the \textit{visited} slots and profiles only the slots that were not visited in earlier passes, until all slots have been visited at least once.
The system uses a high frequency at this stage as it does not have any prior knowledge of the event distribution and thus infrequent wake ups may miss events. 
This is, however, energy demanding. 
Hence, at the beginning of each slot, if the system has enough energy to wake up at a high frequency, it does so to accurately profile the slot by catching all the events that occur in that slot. 
On the other hand, when the stored energy is not enough for high-frequency wake ups, the system remains asleep (i.e., it skips profiling the current slot) and continues to harvest energy, so that it can accumulate energy for waking up at later slots. 
Depending on the application and the expected accuracy of profiling, it might be necessary to repeat the profiling process multiple times until the variance of the number of events caught at each slot reduces, and the process converges.  
The profile approach described above requires multiple passes over the time period to complete profiling all the slots.


\subsection{Phase-2 -- Reinforcement Learning} 
\label{sec:reinforcement_learning}

The goal of reinforcement learning is to determine the wake-up frequency, given the event profile and the harvested energy. Considering the resource-constraints of the system, we adopt a light-weight, table-driven, reinforcement learning technique known as \emph{Q-learning}~\cite{harmon1997reinforcement,qlearning}, which is an effective way of making optimal decisions to achieve the highest cumulative \emph{reward} based on past observations. Q-learning has been adopted by energy harvesting systems that learns the optimal duty-cycle, given the harvested energy~\cite{rlman,roy2018rl,hsu2009reinforcement,shresthamali2017,ACES_workshop,ACES}. In contrast, we consider both the harvested energy as well as the event pattern. In this section, we describe the standard Q-learning approach, followed by an efficient and modular approach that we implement in \Sys.

$\bullet$ \textit{Standard Q-Learning}: Given a system having a set of states and state transition probabilities, Q-learning learns the optimal action at each state such that the total reward is maximized in the long run. The learned knowledge is stored in a table called the \emph{Q-table} which contains a learned value for each state-action pair. We use $S=\{s_1,s_2, \dots,s_n\}$ to denote the set of states and $A=\{a_1,a_2,\dots,a_m\}$ to denote the set of actions. The values in the Q-table are updated iteratively. The general update equation is as follows:
\begin{align}
 \mathrm{Q_{ij} =  Q_{ij} + \alpha[R_{ij}+\gamma \mathop{max}_{k=1,\dots,m}(Q_{i+1,k})-Q_{ij}]} 
\end{align}
where $Q_{ij}$ denotes the value of the state-action pair $(s_i,a_j)$; $R_{ij}$ denotes the reward in state $s_i$ for taking action $a_j$; $\alpha$ denotes the learning rate; $\gamma$ denotes the discount rate; 

When the values in the Q-table converge, the learning process completes. The learner then uses the values stored in the Q-table to choose the optimal action at each state, i.e., selecting the action for which $Q_{ij}$ is the maximum.  

$\bullet$ \textit{Limitation of Standard Q-Learning}: A straightforward application of Q-learning to solve the just-in-time event detection problem would be to define a state space that consists of all possible combinations of energy levels and event arrival probabilities; and the set of actions would be the set of wake-up frequencies that the system supports. This formulation will have a gigantic state space -- which would take forever for the Q-learning algorithm to converge.

$\bullet$ \textit{Q-Learning in \Sys}: \Sys\ exploits the repeated subproblems to break-down the learning problem into multiple, small-sized, quick-converging Q-learning problems. Recall that the event arrival pattern is decomposed in Phase-1 into a set of shapes of different types. Examples of which are shown in Fig.~\ref{fig:Algo_FourEvetTypes}. In \Sys, we apply reinforcement learning for each of these shapes. Hence, there is exactly one Q-table for each shape in \Sys. This modular formulation allows us to reuse the learned Q-table for a certain shape even if it appears many times in the overall event arrival pattern. Furthermore, a separate Q-table per shape allows us to use different learning parameters that are fine-tuned for that shape.  

\begin{figure}[!tb]
    \setlength{\belowcaptionskip}{4pt}
    \centering
    \subfloat[($\text{E}_3,\text{Step}_1$) is first learned.] 
    {
        \includegraphics[width = 0.22\textwidth]{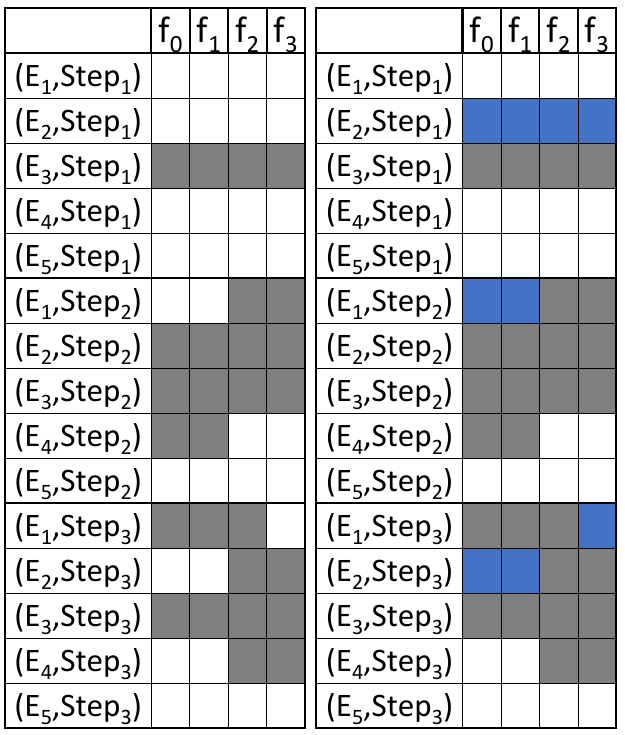}
    }
    \subfloat[($\text{E}_2,\text{Step}_1$) is first learned.]
    {
        \includegraphics[width = 0.22\textwidth]{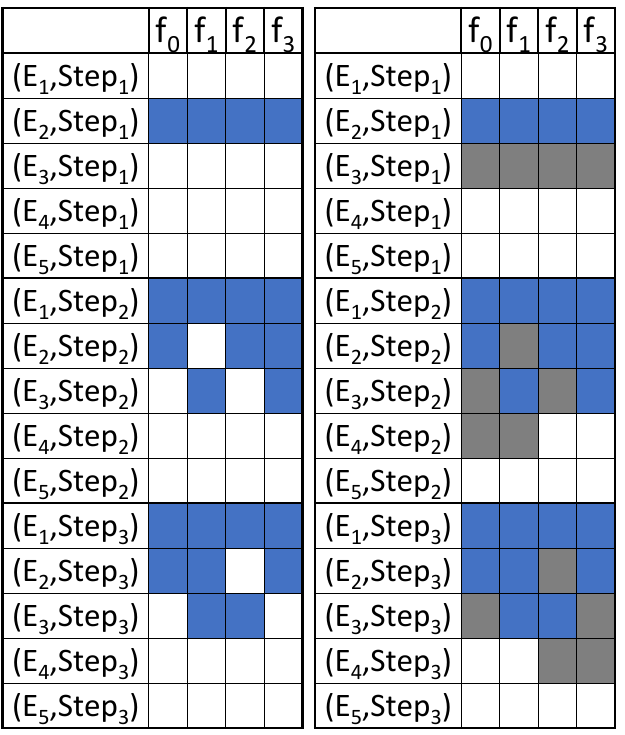}
    }
    \caption{The effect of learning order. Black blocks are learned by ($\text{E}_3,\text{Step}_1$), and blue ones are learned by ($\text{E}_2,\text{Step}_1$).} 
    \label{fig:algo_qtable} 
\end{figure}

We assume that the harvested energy is quantized into $K$ energy levels $\{E_1,E_2,\dots,E_K\}$, a shape spans over $T$ time slots or $T$ steps $\{step_1,step_2,\dots,step_T\}$, and the system supports $N$ different wake up frequencies $\{f_1,f_2,\dots f_N\}$. To formulate the Q-learning problem, we define a state as an (energy, step) pair, and each frequency $f_i$ represents an action. Hence, the Q-table contains $K \times T$ rows and $N$ columns. Fig. \ref{fig:algo_qtable} shows example Q-tables for $K=5$, $T=3$, and $N=4$ having 15 rows and 4 columns.

\vspace{-5pt}
$\bullet$ \textit{Convergence}: In typical Q-learning, a learner can apply the learned knowledge (i.e., begin \emph{exploitation}) only after the whole Q-table has converged. On the contrary, \Sys's Q-learning formulation allows the system to enter the exploitation phase, when only a portion of the Q-table has converged. This is because of the design of the state variable which includes the energy level as part of its definition. More specifically, the level of harvested energy, $E_h$ at the onset of event arrivals (i.e., states of the form $(E_h, step_1)$) limits the energy levels of the next few slots as well as the maximum wake up frequency of those slots. Hence, only a fraction of the Q-Table entries can ever be updated, given the energy level at the entry of a shape, $E_h$. 
In the future, if the energy level at the entry of the same shape changes, although a different set of entries get updated, the original entries still being valid, the learner has to learn less number of entries, and thus converges even faster.

This benefit is illustrated in Fig.~\ref{fig:algo_qtable}. In Fig.~\ref{fig:algo_qtable}(a), a shape is at first entered through the state ($E_3,Step_1$). Later, when the same shape is entered through the state ($E_2,Step_1$), it has less space to explore and will thus converge faster. Fig.~\ref{fig:algo_qtable}(b) shows the opposite scenario when ($E_2,Step_1$) is learned first.   

\begin{algorithm}[t]
\small
 \While{not converged}{
     \For{ $step\leftarrow 1$ \KwTo $EventPeakLength$}{
     \uIf{step == 1}{
         find the righ Qtable based on event shape \;
       }
       $state_{cur}$ = getState($EnergyLevel$, $step$)\;
        $a$ = random()\;   
     execute action $a$\;
     calculate reward $R_{ij}$\;
      $state_{next}$ = getState($EnergyLevel$, $step+1$)\;
     $Q_{ij} = (1-\alpha)Q_{ij} + \alpha(R_{ij}+\gamma \cdot argmaxQ( state_{next},\ :\ ))$
    }
 }
 \caption{Reinforcement Learning}
 \label{algo:algorithm2}
\end{algorithm}
\setlength{\textfloatsep}{0pt}

Algorithm~\ref{algo:algorithm2} shows the pseudocode of the proposed reinforcement learning algorithm in Phase-2. As mentioned before, Phase-1 only tells the system when to wake up. With that information, the system comes to the reinforcement learning model querying for a proper wake-up frequency. 
Algorithm \ref{algo:algorithm2} shows the workflow for one event peak. There may be several event peaks during the entire period as shown in Fig. \ref{fig:varying-charging-capacity}. 
For each peak, the system first identifies the event shape based on the number of step the peak covers, and the composition order of high, $H$ and low, $L$ probability of events in these states, as shown in Fig. \ref{fig:Algo_FourEvetTypes}. Each event shape corresponds to a unique Q-table. If an event shape has no corresponding Q-table, a new Q-table is created. Since we only perform exploration in Phase-2, the action is always picked up randomly. Once an action is chosen and gets executed, the reward is calculated to update the corresponding Q-table. The convergence condition is that the Q-table does not change more than a threshold in a number of consecutive iterations. Once the reinforcement learning model converges, the system moves to Phase-3.

\subsection{Phase-3 -- Exploiting and Probing}
\label{section:adaptive}

$\bullet$ \textit{Exploiting:} Exploiting refers to the act of using the learned knowledge from the earlier phases (i.e., profiling and reinforcement learning) to wake the system up at the right time and at the right frequency.

$\bullet$ \textit{Probing}: Probing refers to the act of detecting changes in the event distribution. Once the system enters Phase-3 and continues to exploit the learned knowledge, it is expected to remain in this phase and continue to detect events until there is a change in the event pattern.  

To perform probing, the system dedicates a small amount of energy to randomly wake up at a few slots where it should not based on the current event profile. If the event distribution remains unchanged, these random wake ups will not observe any new events. However, if events are caught during probing, it could mean that the event distribution may have changed, and thus the system needs to switch to Phase-1 to re-profile the event distribution. Probing enables the system to discover changes in the event pattern and forces the system to adapt to long-term and short-term changes.

$\bullet$ \textit{Long-Term Adaptability}: Long-term adaptability corresponds to the change in event distribution which happens in two ways. First, the event pattern may shift in time. For example, wild animals may come to drink water at noon during the winter, but in the afternoon during the summer. Second, the shapes that constitute the event pattern may change. For example, the new shape may expand or contract in time, or change to a different shape. For example, the event type in Fig.~\ref{fig:Algo_FourEvetTypes}(a) may expand and become like Fig.~\ref{fig:Algo_FourEvetTypes}(b).
Phase-1 is responsible for adapting the system to these changes.

$\bullet$ \textit{Short-Term Adaptability}: Short-term adaptability corresponds to the system entering the event shape at different entry energy states. As shown in Fig.~\ref{fig:optimalwakinguppolicy}, even for the same event shape, if the system enters the shape at different energy levels, the optimal wake-up policy is different. The reinforcement learning model of Phase-2 is responsible for adapting the system to these changes. Notice that once the system learns how to adapt, given a particular shape and the entry energy level, the knowledge is permanent and the system does not need to learn it again and that is why we call it short-term adaptability.

\section{Experimental Setup}

\begin{figure}[t] 
    \setlength{\belowcaptionskip}{-18pt}
    \centering
    \subfloat[Image-based.]{
        \includegraphics[width = 0.19\textwidth]{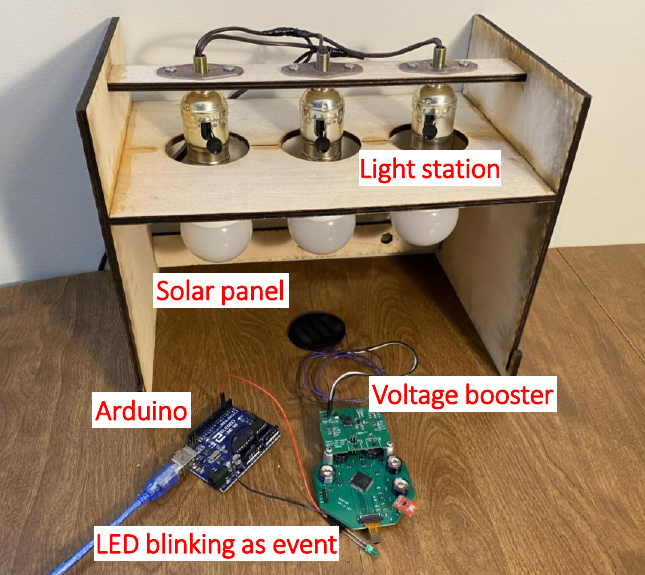} 
    }
    \subfloat[Audio-based.]{
        \includegraphics[width = 0.223\textwidth]{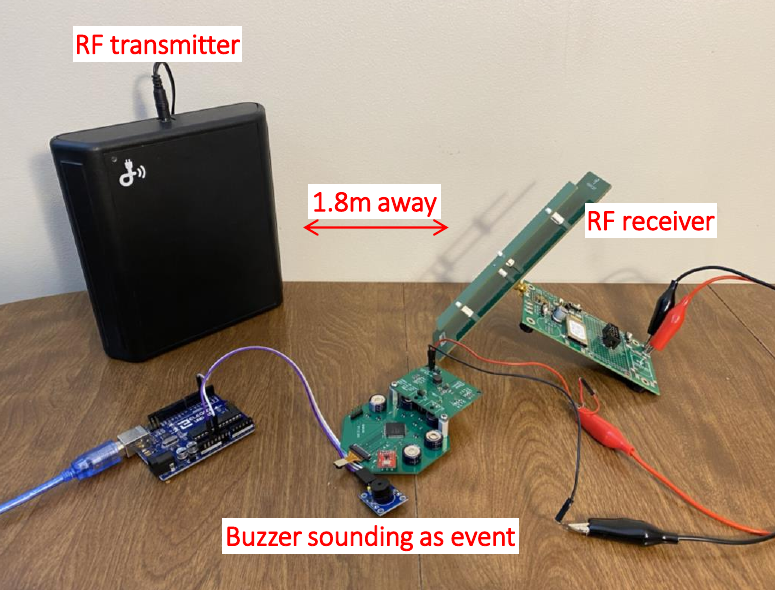}     
    }
    \caption{Hardware experiment setup.}
    \label{fig:hardware_real}
\end{figure}

We conduct extensive evaluation of \Sys using both real systems as well as simulations. 

\subsection{Real Systems}
\vspace{-4pt}
We implement two real systems that use a camera and a microphone sensor, respectively. The hardware platforms are shown in Fig.~\ref{fig:hardware_real}.

$\bullet$ \textit{Image-based System}: The image-based system uses an ultra low power camera sensor, HM01B0 (consuming $<$2mW at QVGA, 30FPS), to capture images at 1 frame/sec and uses the on-board MSP430FR5994 MCU for image processing and event detection. To emulate an event, we blink an LED (controlled by an Arduino) at certain times following event distribution patterns such as the ones shown in Fig.~\ref{fig:Algo_FourEvetTypes}. The MCU uses a threshold-based light pattern recognizer~\cite{zygarde} to detect the presence of an event. We referred ~\cite{josephson2019wireless} for implementing our image-based experiments.     

The system is powered by harvesting light energy from a rig that consists of an array of bulbs whose intensity is controlled to simulate the varying intensity of sunlight in a real-world scenario. The \Sys's capacitor array uses two 12mF, two 47mF and a 110mF capacitors to store the harvested energy.

$\bullet$ {\textit{Audio-based System}}: The audio-based system uses a low power microphone, ADMP401, to capture audio at 8KHz and uses the same MCU for audio processing and event detection. Similar to the image-based system, to emulate an event, we sound a buzzer (controlled by an Arduino). The MCU performs FFT and detects the presence or absence of the buzzer sound which occurs at 2KHz.

The system is powered by harvesting RF energy. We use a TX91501 powercaster transmitter and a P2110 powerharvester receiver. The RF energy harvesting pattern is controlled by changing the relative distance between the transmitter and the receiver to emulate real-world scenarios. The \Sys's capacitor array uses a 4.7mF, two 12mF, and one 47mF capacitors to store the harvested energy. Note that this setup uses a smaller capacity array than the image-based systems since the audio-based system requires relatively less energy.

\subsection{Learning Algorithm Parameters}
\vspace{-4pt}
In both systems, the learning rate, $\alpha$ = 0.7 and discount rate, $\gamma$ = 0.618. The wake up frequencies for different actions are: $f_3$ = 1Hz, $f_2$ = 0.5Hz, $f_1$=0.2Hz, and $f_0$=0Hz (never awake). The reward is +10 for catching an event and -1 for nothing. The unit of time is 1 second and each state is 30 seconds long. The period is set to 20 minutes so that the system can complete multiple iterations of learning in a reasonable amount of experiment time. The maximum duration of an event peak is set to 2 minutes (i.e., 10\% of the period).

\subsection{Evaluation Metrics}
\vspace{-4pt}
We define two evaluation metrics to quantify the performance of event detection:

$\bullet$\textit{Total Catches:} This refers to the total number of time units when the system woke up and were able to detect an event. In other words, irrespective of the number of times a system wakes up, if it detects $n$ events, then Total Catch = $n$.

$\bullet$ \textit{Energy Efficiency:} This refers to the portion of harvested energy spent by the system (e.g., to wake up and detect events) when there is actually an event. Harvested energy is wasted when the system wakes up but there is no event to detect. An energy efficiency of 100\% refers to an ideal system which only wakes up when events are happening.

\subsection{Baseline Solutions}
\vspace{-4pt}
We compare \Sys against three other systems:

$\bullet$ \textit{GT} refers to ground truth system that catches all events.

$\bullet$ \textit{CTID} represents existing energy harvesting systems~\cite{ACES,ACES_workshop,rlman,roy2018rl} that wake up after the capacitor is charged to the turn-on threshold and immediately consume the energy until the system reaches the turn-off threshold. This also include systems that dynamically change the duty cycle only based on the remaining energy~\cite{dewdrop,tragedy,capybara}.

$\bullet$ \textit{CTIDpro} represents a CTID system that implements the event profiling scheme (similar to \Sys's Phase-1) but does not implement reinforcement learning (i.e., \Sys's Phase-2). We include CTIDpro to demonstrate the necessity and effectiveness of \Sys's reinforcement learning.

\begin{figure}[!t] 
    \setlength{\belowcaptionskip}{-18pt}
    \centering
    \subfloat[Phase 1] 
    {
        \includegraphics[width = 0.22\textwidth]{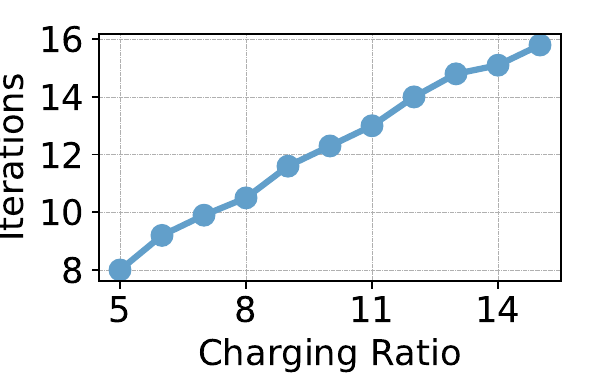}
    }
    \subfloat[Phase 2]
    {
        \includegraphics[width = 0.22\textwidth]{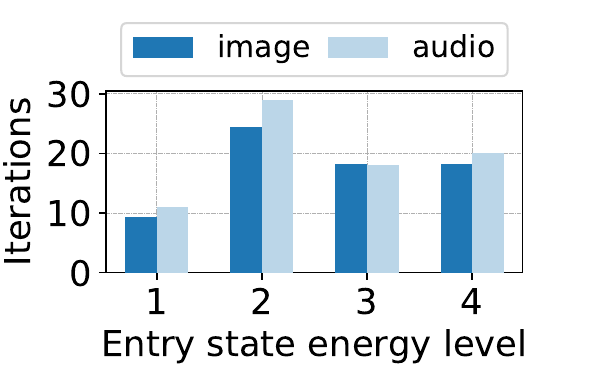}
    }
    \caption{Iterations to converge (a) Phase-1, and (b) Phase-2.} 
    \label{fig:eval_ConvRatePhase2} 
\end{figure}

\begin{figure*}[t] 
    \setlength{\belowcaptionskip}{-16pt}
    \centering
    \subfloat[Image-based System] 
    {
        \includegraphics[width = 0.48\textwidth]{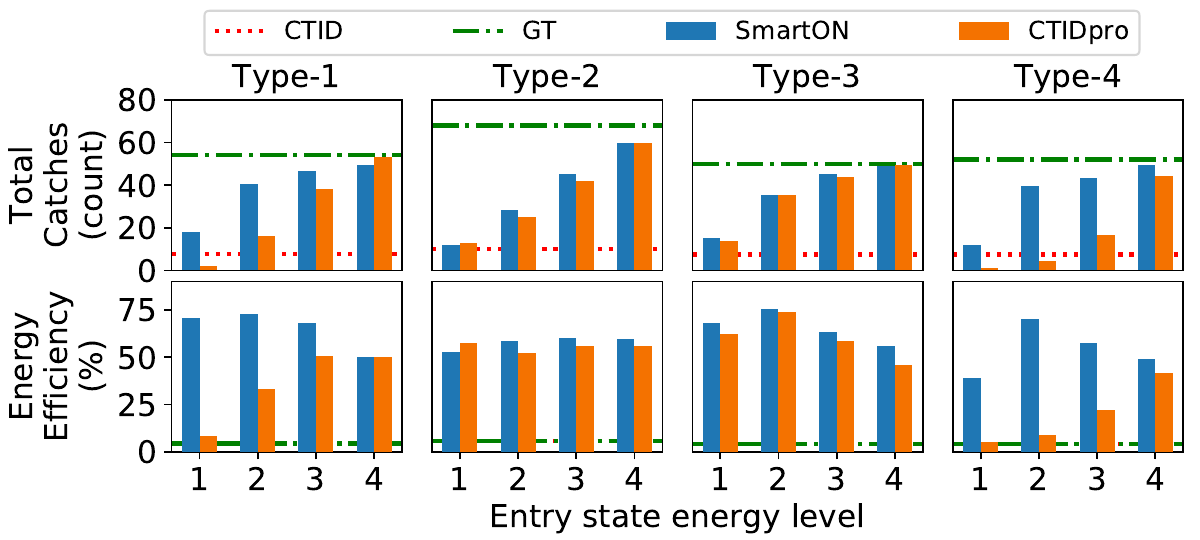}
    }
    \subfloat[Audio-based System]
    {
        \includegraphics[width = 0.48\textwidth]{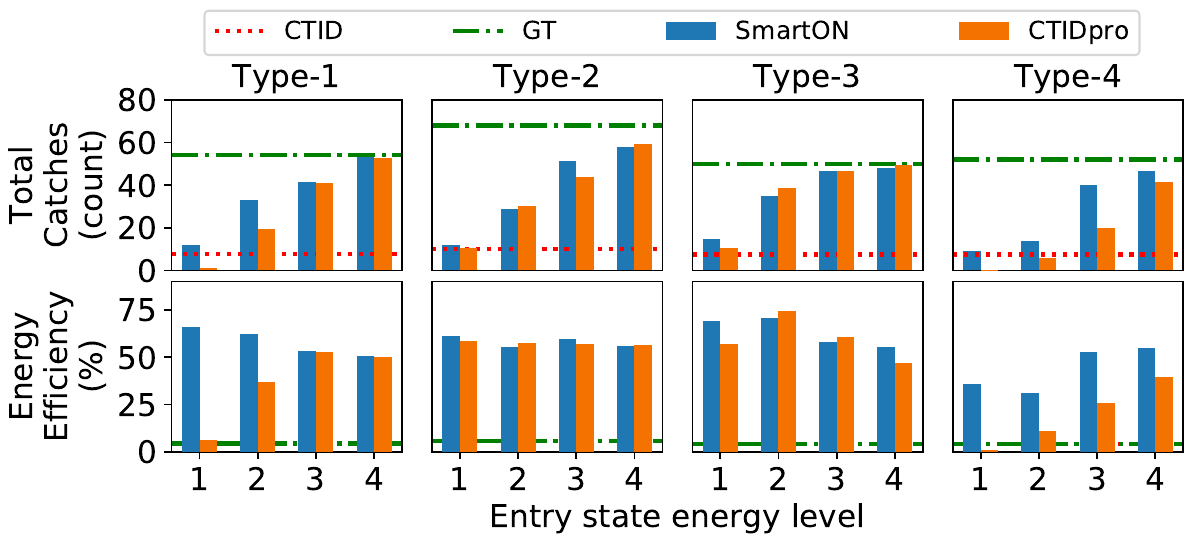}
    }
    \caption{Performance of event detection for four event types. The parameters for generating event distributions are partially based on ~\cite{xue2018activity}.}
    \label{fig:eval_Perf}
\end{figure*}


\section{Experimental Results}

We describe the convergence of the first two phases of \Sys and the performance of event detection after the system is converged.

\subsection{Phase-1 Convergence}
\vspace{-4pt}

The Phase-1 of \Sys converges when all but a few states in the beginning have been explored via high-frequency wake-ups. We ignore the first few states as these states may never converge as there is little energy at the beginning. We measure the convergence rate in terms of the number of iterations the algorithm takes to converge. 

Convergence of \Sys's Phase-1 depends on the intensity of the energy source, which is characterized by the capacitor's charging to discharging ratio, or simply the \emph{Charging Ratio}. A charging rate of $r$ means the system has to harvest energy for $r$ units of time to be able to wake up for one unit of time.

Fig.~\ref{fig:eval_ConvRatePhase2}(a) shows that the convergence rate of Phase-1 linearly increases with the charging ratio. Note that the charging ratio is application-dependent. The two systems we implemented require a charging ratio of around 9.    

\subsection{Phase-2 Convergence}
\vspace{-4pt}
The Phase-2 of \Sys converges when the entries of the Q-table does not change or the change is below a small threshold. Since the Q-table in \Sys is partitioned by the energy level of the entry state, each of these smaller-sized Q-learning problems converges separately. We measure the convergence rate in terms of the number of iterations they take to converge.

Convergence of \Sys's Phase-2 depends on the size of the state space, i.e., the number of energy levels and the number of steps that the event peak covers. Because of the partitioning scheme in \Sys, some of the partitions that converge early reduces the converge time for the yet-to-converge partitions as they often share common entries in the Q-table that may have already been explored.

Fig.~\ref{fig:eval_ConvRatePhase2}(b) shows the number of iterations needed for different levels of entry state energy. We observe that the middle-energy level entry states need more iterations to converge as they may transition to both higher and lower energy levels, and thus have a larger overlap with others in the Q-table. Note that we have used only four levels in this evaluation, which is sufficient for the real systems we developed. More results on the convergence for up to 10 levels are shown in the Appendix.

\begin{figure}[t] 
    \setlength{\belowcaptionskip}{-15pt}
    \centering
    \includegraphics[width = 0.46\textwidth]{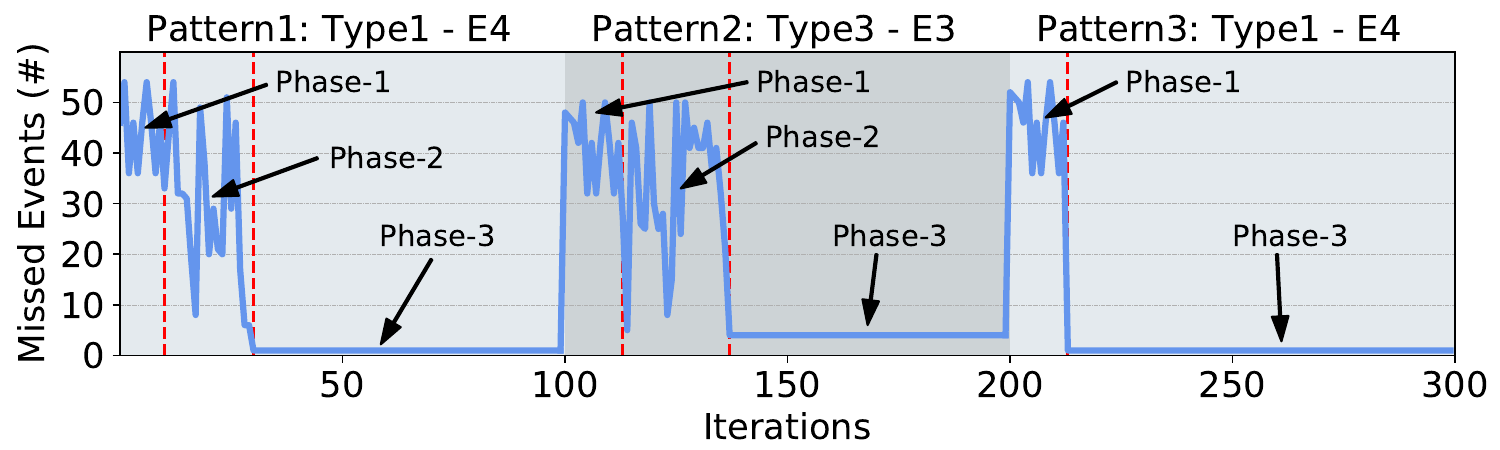} 
    \caption{Performance during adaptation. There are three patterns shown in this figure. E denotes entry state energy level. }
    \label{fig:eval_real_adaptation}
\end{figure}

\subsection{Effect of Event Types on Event Detection}

During its operation, a system encounters a number of event peaks having different shapes and duration (e.g., Fig.~\ref{fig:Algo_FourEvetTypes} shows four example event types). The performance of event detection not only depends on the event type, but also on the level of harvested energy when the system enters a peak.

In this experiment, We measure the total catches and energy efficiency of \Sys and three baselines. 
Fig.~\ref{fig:eval_Perf} shows the mean performance numbers across all event types for each of the four levels of entry state energy.
We observe that the results for image- and audio-based systems are consistent. The GT and the CTID capture the highest and the lowest number of events, respectively. CTIDpro and \Sys catch more events when they enter an event peak with higher levels of energy as they have more energy to wake up when energy is high. 
The energy efficiency of both GT and CTID are very poor. These systems waste over 95\% energy by waking up when there is no event to capture. 

We observe that CTIDpro only performs similar to \Sys for Type-2 and Type-3 events because for those two types most events are likely to occur at earlier steps and thus the system does not have to conserve energy for later steps. For Type-1 and Type-4 events, CTIDpro for lack of reinforcement learning fails to allocate energy for later steps and performs poorly when the entry state energy level is low. This shows the effectiveness and necessity of the reinforcement learning module in \Sys.

Overall, on average, \Sys\ captures 1X-7X more events and is 8X-17X more energy-efficient than existing energy harvesting systems which employ a CTID strategy.

\subsection{Adaptation to New Patterns}
\Sys\ can adapt itself when event pattern changes. We show how \Sys\ performs during the adaptation process in Fig.~\ref{fig:eval_real_adaptation}. 

The system first enters event Type-1 with an energy level of 4. It stays in Phase-1 to profile the event distribution for around 10 iterations and then in Phase-2 for another 20 iterations to learn the optimal wake-up strategy. We can see that the missed events are dramatically decreased after it enters Phase-3. 
Next, the event pattern changes into Type-3 with an energy level of 3. \Sys\ follows the same workflow as the previous pattern. 
Finally, it comes to the third pattern, Type-1 with an energy level of 4, which is the same pattern as the first one. This time, \Sys does not need to enter Phase-2 as the wake-up strategy has already been learned, so we can see that it converges faster than the first one as Phase-2 is skipped.

\subsection{Simulation results}
Simulation are developed based on a python package called \textit{gym} which is a toolkit for developing and comparing reinforcement learning algorithms. The simulation has the same time period as the real-system experiments in order to make their results consistent. The purpose of simulation is to investigate more parameters and to show our algorithm produces consistent results in repeated runs. 
In order to better demonstrate the benefit of our quick-converging Q-learning scheme, we use the simulation with 10 energy levels, which means there are 10 entry states.

\begin{figure}[t] 
    \setlength{\belowcaptionskip}{-10pt}
    \centering
    \includegraphics[width = 0.45\textwidth]{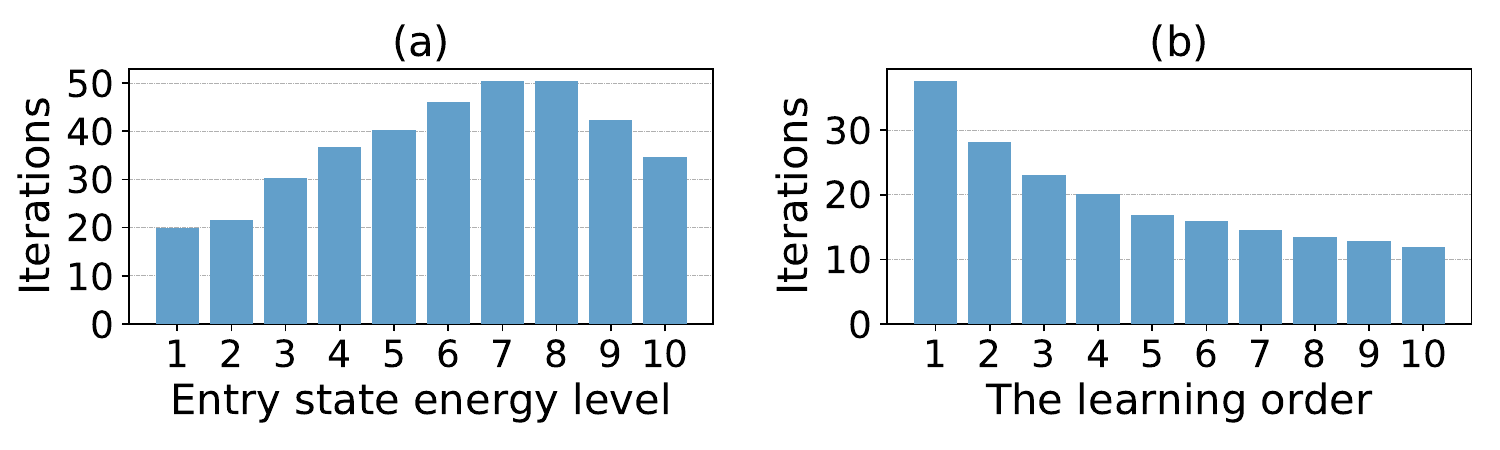} 
    \caption{Convergence rate of Phase-2 - simulation results.}
    \label{fig:eval_simu_ConvRatePhase2}
\end{figure}

\textbf{Convergence rate}. Fig.~\ref{fig:eval_simu_ConvRatePhase2}(a) shows the convergence rate for each entry state when they are the first to be learned. We can see that allowing each entry state to converge individually can greatly speed up the exploitation, which requires 37 iterations on average, compared to the standard Q-learning algorithm which would not start exploitation until the entire Q-table is converged, which requires 370 iterations in total.   
In Fig.~\ref{fig:eval_simu_ConvRatePhase2}(b), we shuffle the learning order of the ten entry states in each run and then calculate the average result for each order. We can clearly see that later learned entry states need less iterations to converge as some of the common space may have already been explored by the entry states that are learned earlier. This result supports the explanation of Fig.~\ref{fig:algo_qtable}. 

\begin{figure}[t] 
    \centering
    \includegraphics[width = 0.49\textwidth]{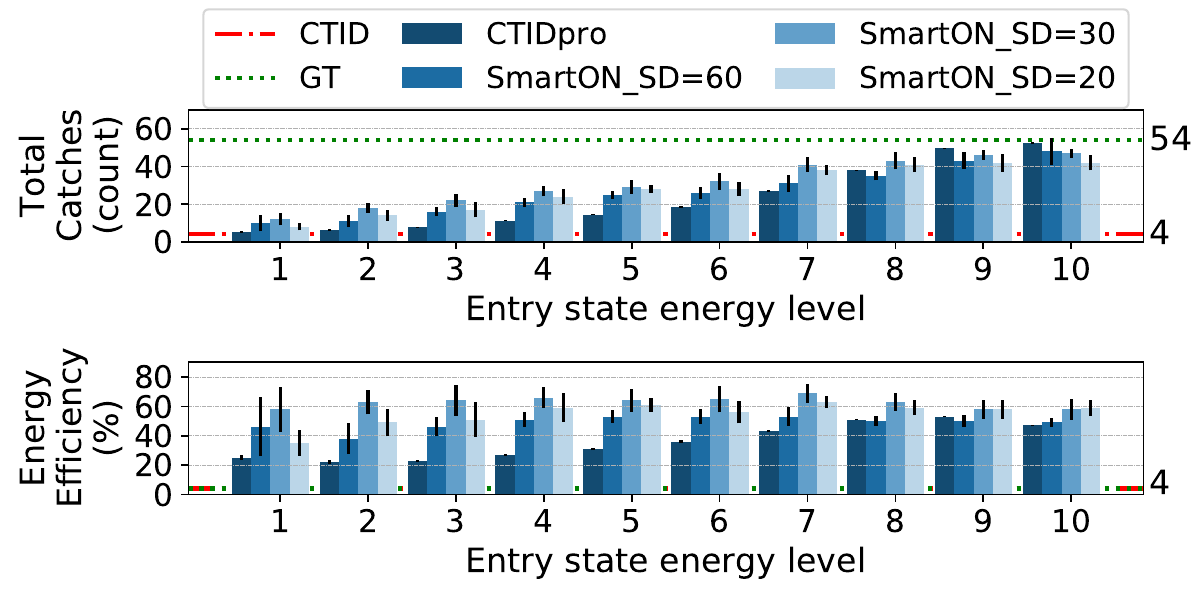} 
    \caption{Effects of state duration - simulation results (Type-1). We set the state duration to 20, 30 and 60 seconds, respectively. }
    \label{fig:eval_simu_PerfEntryState}
\end{figure}

\begin{figure}[t] 
    \centering
    \includegraphics[width = 0.45\textwidth]{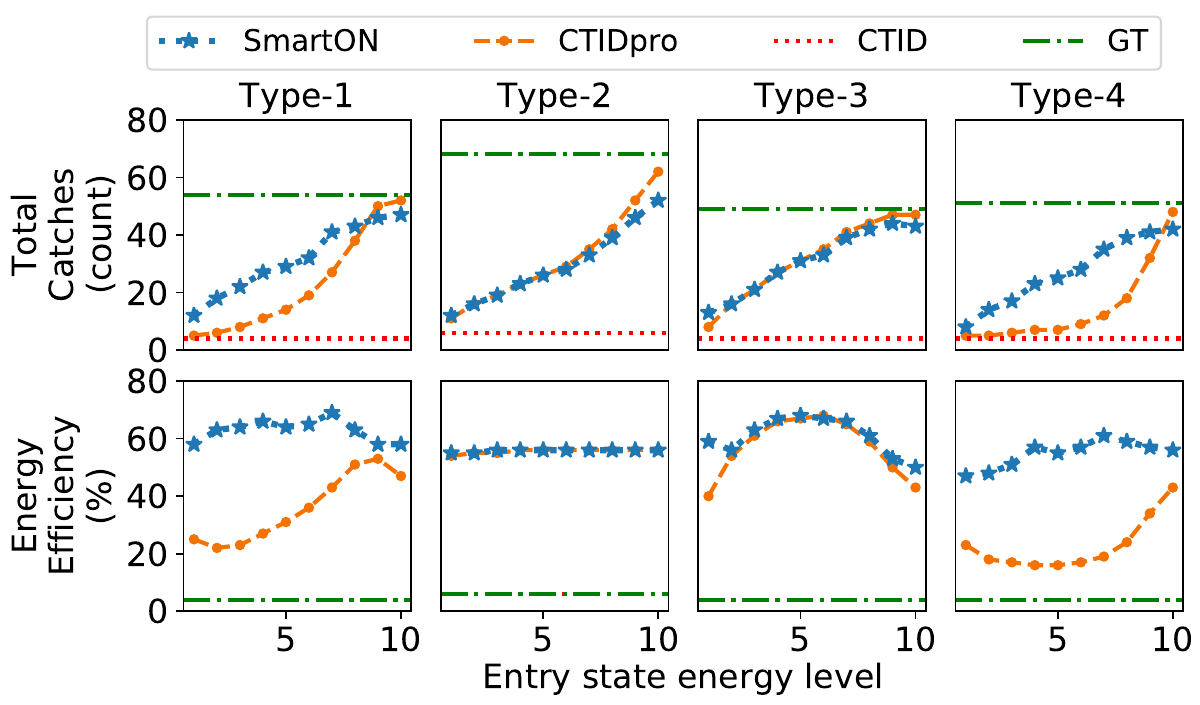} 
    \caption{Effects of event type - simulation results.}
    \label{fig:eval_simu_PerfEventType}
\end{figure}

\textbf{Detection performance}. Fig.~\ref{fig:eval_simu_PerfEntryState} shows the effects of state duration. State duration means how long each state lasts. We can see that the duration of 30 seconds yields best results based on which we use 30-second duration in the hardware experiments. The error bars also show that our algorithm is stable in producing consistent results in different runs. Fig.~\ref{fig:eval_simu_PerfEventType} shows the result of four event types. The conclusion we can draw here, with regard to the advantage of \Sys\ over other baselines, is consistent with that of our hardware experiments.

\subsection{System Overhead}
We analyse the system overhead in terms of memory, execution time and energy consumption. 
We use the Analog Discovery to measure the execution time and power consumption.   
The CTIDpro system has similar overhead with a CTID system so we only mention CTID in the following comparison.

\subsubsection{\textbf{Memory}}
MSP430FR5994 has two types of memory, 256KB of FRAM and 8KB of RAM. The CTID system consumes 8.9KB of FRAM and 1.3KB of RAM. Compared to the CTID system, \Sys\ needs extra 10KB of FRAM for the reinforcement learning algorithm, and 0.5$\sim$2KB of FRAM for saving each qtable. The total amount of FRAM needed for qtables depends on how many event shapes the system has encountered so far.

\subsubsection{\textbf{Execution time}}
Fig. \ref{fig:eval_real_Overhead}(a) shows the execution time.
The CTID system does not include the reinforcement learning  module.
The time overhead caused by reinforcement learning is very small, only 103ms. The insignificance of the overhead caused by reinforcement learning is even more obvious in Fig. \ref{fig:eval_real_Overhead}(b) in which the energy consumption of reinforcement learning only accounts for 24.2\% of that of one image-based event detection given that one iteration may have dozens or even hundreds of times of event detection. 
\subsubsection{\textbf{Energy consumption}}
\begin{figure}[t] 
    \setlength{\belowcaptionskip}{-15pt}
    \centering
    \includegraphics[width = 0.5\textwidth]{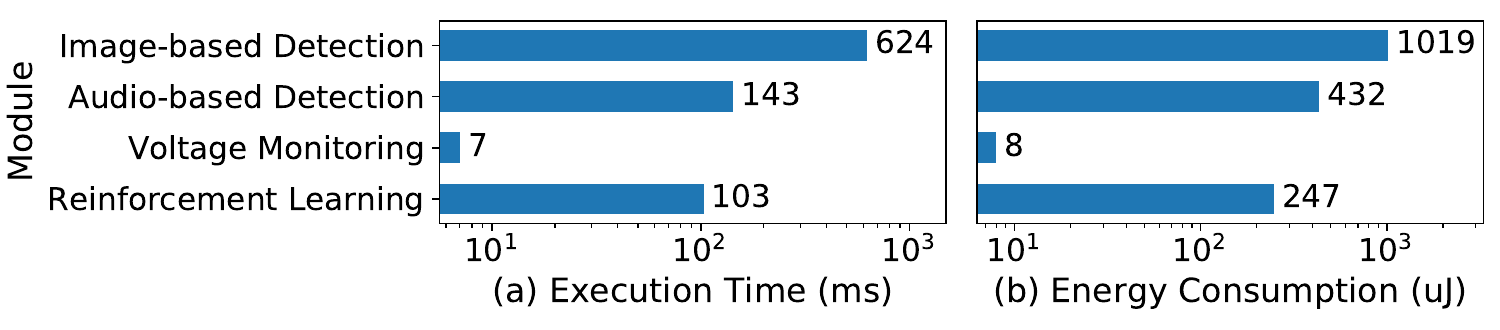} 
    \caption{Overhead (time execution and energy consumption). The overhead of event detection includes sensing and classification, and only corresponds to one single detection. The overhead of reinforcement learning is for one iteration.} 
    \label{fig:eval_real_Overhead} 
\end{figure}

Fig. \ref{fig:eval_real_Overhead}(b) shows the energy consumption.
Image-based event detection consumes 2.3X times of energy as the audio-based event detection. 
The extra hardware components of \Sys\ also cause energy overhead. \Sys\ has several switches that turn on or off capacitors on-the-fly. Each such switch has one resistor and two MOSFETs which cause extra energy consumption.
The static current consumption of each such switch unit amounts to 2.4$\mu$A. Besides, in real-life scenarios, \Sys\ would need an external RTC (e.g. Botoks\cite{botoks}) which is not implemented in our experiments. We can refer the data from Botoks and its static current consumption of four-tier RTC amounts to 0.958$\mu$A. With all of that combined, compared to a CTID system, the extra power overhead of \Sys\ with four switches is around 10.6$\mu$W.

\section{Limitation and future work}
\vspace{-5pt}

Despite the promising results in just-in-time active event detection, \Sys has a few limitations. 

First, \Sys learns to wake up at the right time and at the right frequency only when the event distribution pattern is learnable and not random. Furthermore, if the pattern changes too fast, \Sys may not perform as well as expected. 

Second, the current implementation of \Sys does not consider the relationship between consecutive event peaks. \Sys learns the wake-up policy for each event peak individually. The performance of \Sys can be improved further if an optimal wake-up policy is learned jointly for consecutive event peaks.

Third, \Sys implements a multiphase learning framework where learning happens in Phases-1 and Phase-2, and the application of the learned knowledge happens in Phase-3. While this separation makes the system design simpler, by allowing Phase-3 to be executed in parallel with Phases-1 and Phase-2, the responsiveness of \Sys when dealing with varying event patterns can be improved further.

\section{Related Work}

Works related to passive event detection are summarized in \cite{contexttrigger,campbell2016,campbell2014energy} and include door-opening~\cite{campbell2014energy}, airflow~\cite{trinity2013,trinity2018}, water usage~\cite{doubledip,campbell2014water}, occupancy~\cite{campbell2014energy}, power usage~\cite{monjolo2013}, human-walking~\cite{stepcounter} detection. As mentioned in \cite{contexttrigger}, all these works have a common drawback that the event itself must be able to captured by the harvester to produce a trigger signal. However, not every physical parameter change can trigger the harvester and that is why active event detection is necessary. 

Prediction-based approaches~\cite{moser2009prediction,kansal2007prediction,sudevalayam2010predictionsurvey,buchli2014dynamic} are used to predict the availability of energy in the future 
and the system uses this information to change the duty cycling on-the-fly. Reinforcement learning based approaches~\cite{rlman,roy2018rl,hsu2009reinforcement,shresthamali2017,ACES_workshop,ACES} only use the energy residue, along with other information, to form the state space of the reinforcement learning algorithm. As they focus on wireless sensor networks where the continuity of sensing is their main concern, they do not learn from the perspective of the event pattern and thus not applicable to active event detection that we are dealing with, where events and harvestable energy are not aligned in time.
Though the work~\cite{ACES} claims that it is generalized to event-driven sensing, as far as we understand it actually does not learn the event pattern at all according to its formation of state and reward.

\section{Conclusion}

A new reinforcement learning-based active event detection system is introduced. The proposed system runs on harvested energy, learns from the perspectives of energy source, and event pattern and adapts its duty cycle to match the distribution of event arrival pattern. The proposed multiphase learning framework enables rapid adaptation to changes in the event pattern.
A dedicated power management module has been proposed which expands the storage capacitance on-the-fly. An image-based and an audio-based application have been implemented and evaluated. Both real-system experiments and software simulations show that the proposed system captures more events and is more energy-efficient than the baselines.

\section{Acknowledgement}

This paper was supported, in part, by NSF grants CNS-1816213 and CNS-2047461.

\bibliographystyle{ieeetr}
\bibliography{SmartON.bib}


\end{document}